\documentclass{iau} 
\usepackage{graphicx}
\usepackage{amsmath}
\usepackage{amssymb}
\usepackage{natbib}
\def\aap{{Astron.~Astrophys.}}                
\def\apj{\textit{ApJ}}
\def\apjl{\textit{ApJL}}
\def\apjs{\textit{ApJS}}
\def\mnras{\textit{MNRAS}}
\def\prc{\textit{Phys.~Rev.~C}}        
\def\pasa{\textit{PASA}}               
\def\pasj{\textit{PASJ}}               
\def\physscr{\textit{Phys.~Scr.}}   
\def\nat{\textit{Nature}}

\title{The Core-Collapse Supernova Explosion Mechanism}

\author[Bernhard M\"uller]   
{Bernhard M\"uller$^{1,2}$}

\affiliation{$^1$Astrophysics Research Centre, Queen's University Belfast\\BT7 1NN, Belfast, Northern Ireland \\ email: {\tt b.mueller@qub.ac.uk} \\[\affilskip]
$^2$Monash Centre for Astrophysics, Monash University \\ Clayton, VIC 3800, Australia}

\pubyear{2017}
\volume{329}  
\jname{The Lives and Death-Throes of Massive Stars}
\editors{A.C. Editor, B.D. Editor \& C.E. Editor, eds.}
\begin{document}

\maketitle

\begin{abstract}
The explosion mechanism of core-collapse supernovae is a long-standing
problem in stellar astrophysics. We briefly outline the main
contenders for a solution and review recent efforts to model
core-collapse supernova explosions by means of multi-dimensional
simulations. Focusing on the
neutrino-driven mechanism, we summarize currents efforts to predict
supernova explosion and remnant properties. 
\keywords{
Supernovae: general,
stars: evolution, 
stars: interiors,
hydrodynamics,
instabilities,
convection,
neutrinos,
turbulence,
methods: numerical
}
\end{abstract}

\firstsection 
\section{Introduction}
Many open questions still surround the terminal gravitational core
collapse of massive stars.  Theory has by now established a clear
picture of the first two acts of their death throes. The collapse of
the iron (or in some cases O-Ne-Mg) core is triggered by
deleptonization and/or photo-disintegration of heavy nuclei. The
collapse is then halted once the core reaches supranuclear densities,
and its elastic rebound launches
a shock wave that quickly stalls due to energy losses
by nuclear dissociation and neutrino losses. Theory has yet to fully
explain the subsequent acts in the drama -- the revival of the shock
and the development of a supernova explosion -- let alone the full
systematics of supernova explosion and remnant properties.  Among the
proposed explosion scenarios (see \citealt{janka_12} for an overview),
the neutrino-driven mechanism and various flavours of a
magnetorotational mechanism have been most thoroughly explored.

\begin{figure}[b]
\begin{center}
 \includegraphics[width=0.64 \textwidth]{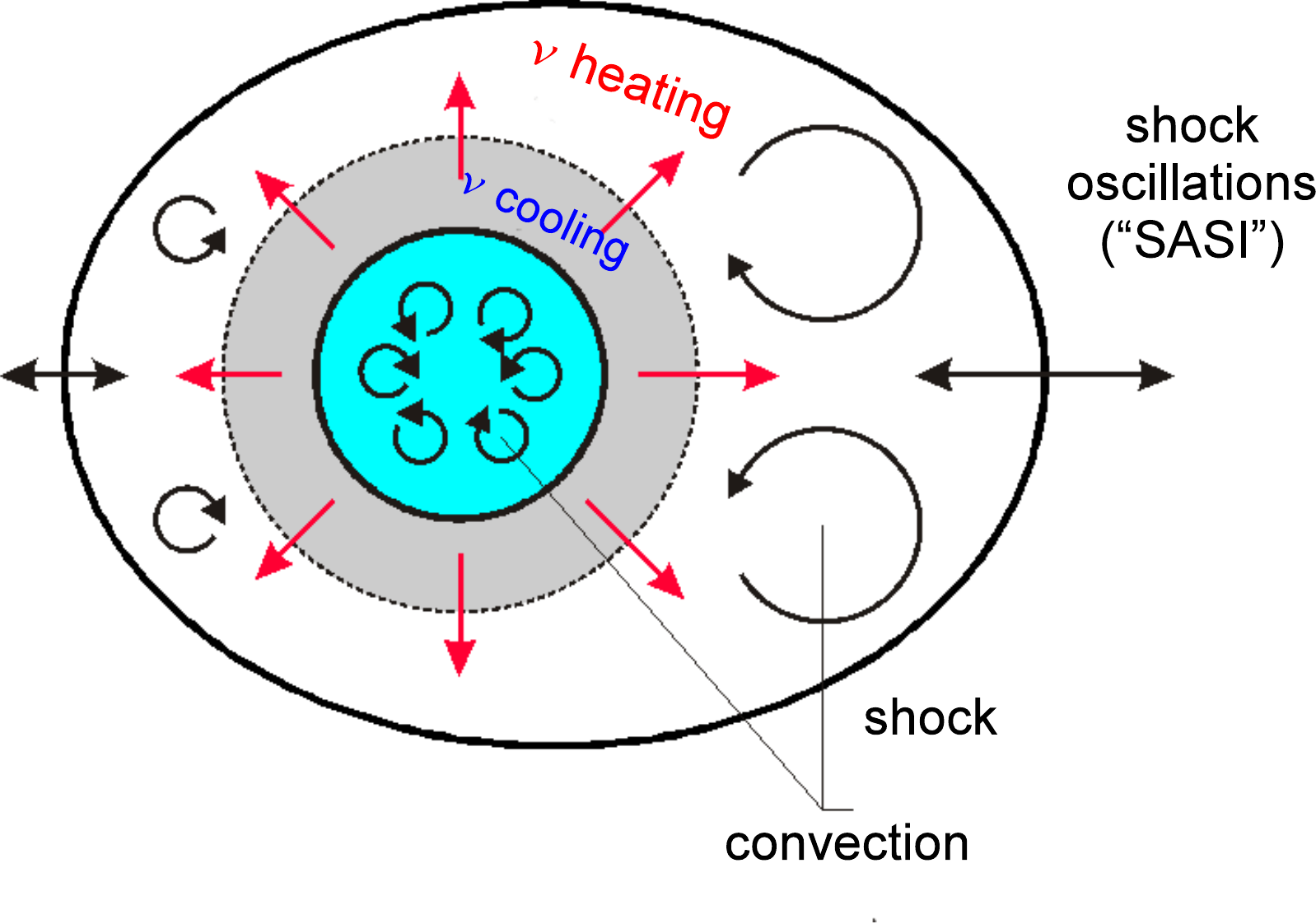} 
 \caption{Sketch of the crucial elements of the neutrino-driven
   explosion mechanism. Neutrinos are emitted from the accretion layer
   (grey) around the PNS and from its 
   core.  A fraction of the electron-flavour neutrinos are reabsorbed
   in the ``gain layer'' behind the shock.  In the gain region,
   non-spherical flow can develop because neutrino heating drives
   convective overturn and because of an advective-acoustic
   instability (SASI) that manifests itself in the form of large-scale oscillatory
   motions of the shock. Energy and lepton number loss also drives
   convection inside the PNS (cyan), which does not
   play a prominent role for shock revival, however.  
   \label{fig:neutrino}}
\end{center}
\end{figure}

Here we shall mostly focus on the neutrino-driven mechanism, which has
the virtue of not requiring special evolutionary
channels for producing progenitors that
spin more rapidly than generically predicted
by current stellar evolution models \citep{heger_05,cantiello_14}
for the bulk of massive stars.
In the neutrino-driven scenario (sketch in
Figure~\ref{fig:neutrino}), shock revival is accomplished thanks
to the increase of the post-shock pressure due to partial
reabsorption of neutrinos that stream out from the young proto-neutron
star (PNS) and the cooling layer of accreted material on its surface. Except
for the low-mass end of the progenitor spectrum (e.g.\
\citealp{kitaura_06,melson_15a}), this mechanism has been found to depend
critically on multi-dimensional (multi-D) instabilities such as buoyancy-driven
convection \citep{herant_94} and the standing accretion shock instability
(SASI, \citealp{blondin_03}) to foster neutrino-driven runaway
shock expansion by a
combination of effects including the mixing of the post-shock region
and the provision of turbulent stresses \citep{murphy_12}. Demonstrating
quantitatively that shock revival can be achieved in this manner is a
notoriously hard problem and requires sophisticated
multi-physics simulations that incorporate neutrino transport,
multi-D fluid flow, general relativity, and the pertinent
microphysics (neutrino interaction rates, nuclear equation of state)
at an appropriate level of accuracy.

\section{Progress and Challenges in Modelling
Neutrino-Driven Explosions} As they progressed in sophistication over
several decades, simulations of neutrino-driven explosion
followed a tortuous path between success and failure. The most recent step to
three-dimensional (3D) multi-group neutrino hydrodynamics simulations
has been no exception. After experiments with simple ``light-bulb''
models in 3D \citep{nordhaus_10,hanke_12,couch_12b}, fully-fledged 3D
three -flavour neutrino hydrodynamics simulations soon became available 
\citep{hanke_13}, and by now there is already a modest number of
successful explosion models for progenitors
from $9.6M_\odot$ to $20 M_\odot$ obtained with rigorous
\citep{melson_15a,melson_15b,lentz_15} or simplified
\citep{takiwaki_14,mueller_15b} energy-dependent neutrino transport.
 By and large, the 3D models have exhibited
a trend towards \emph{missing} \citep{hanke_12,tamborra_14a} or \emph{delayed}
explosions, i.e.\ they show less optimistic conditions for shock
revival compared to the massive corpus of successful axisymmetric (2D)
simulations that has been amassed by different groups
\citep{janka_12b,bruenn_13,nakamura_14,summa_16,oconnor_16}.  The
adverse effect of the third dimension has mainly been attributed to
the different behaviour of turbulence in 3D and 2D (though this is
only the most important element of a more nuanced view as pointed out
by \citealt{mueller_16b} and \citealt{janka_16}). In 2D the inverse
turbulent cascade proves conducive to the emergence of large-scale
modes and allows convection and the SASI to become more vigorous
\citep{hanke_12}.  The heating conditions are not
\emph{substantially} more pessimistic in 3D, however. A quantitative
analysis  reveals that even non-exploding
models, such as the $27 M_\odot$ \citep{hanke_12}, $11.2 M_\odot$, and
$20 M_\odot$ \citep{tamborra_14a} runs of the MPA group, come close to
the critical conditions for neutrino-driven runaway shock expansion.

Nonetheless, the results of recent simulations
pose a
problem. On the one hand, 3D models have not even been able to clearly
establish the efficacy of the neutrino-driven mechanism for
progenitors in the mass range of $10\text{--}15 M_\odot$, where there
is clear observational evidence for ``explodability''
\citep{smartt_15}. This may merely be a result of poor sampling as
only one case in the middle of this range has been simulated with
state-of-the-art transport so far \citep{tamborra_14a}.  On a more
serious note, the delayed onset of explosions in 3D models makes it
more difficult to reach significant explosion energies, as the decline
of the neutrino luminosities and the mass in the gain region reduces
the amount of energy that can be pumped into the ejecta by neutrino
heating. 

The more reluctant development of explosions in 3D has therefore
prompted efforts to identify missing physical ingredients for earlier
and more robust shock revival. The gist behind these ideas can be
understood if the problem of shock revival is phrased in terms of a critical limiting
luminosity $L_\mathrm{\nu,crit}$
as a function of accretion rate $\dot{M}$
 for stationary accretion flow
 in spherical
symmetry \citep{burrows_93}. This concept has been generalized to account for other
parameters of the accretion flow and multi-D effects
\citep{janka_12,mueller_15a,janka_16} in a phenomenological manner and
captures the transition to shock revival remarkably well even in
multi-D \citep{summa_16}. In terms of
the averaged root-mean-square energy $E_\nu$ of $\nu_e$ and
$\bar{\nu}_e$, the accretion rate $\dot{M}$, the PNS
mass $M$, the gain radius $R_\mathrm{g}$, the average net binding
energy $|e_\mathrm{tot,g}|$ in the gain region, and the average
$\langle \mathrm{Ma}^2\rangle$ of the (squared) turbulent Mach number
in the gain region, \citet{summa_16} find
\begin{equation}
\label{eq:lcrit}
(L_\nu E_\nu^2)_\mathrm{crit} 
\propto
\left(M \dot{M} \right)^{3/5}
|e_\mathrm{tot,g}|^{3/5}
R_\mathrm{g}^{-2/5}
\left(1+\frac{4}{3}\langle \mathrm{Ma}^2\rangle \right)^{-3/5}.
\end{equation}
The key to more robust explosions consists in identifying
effects that change the terms in Equation~(\ref{eq:lcrit})
in a favourable direction.

\subsection{Uncertainties in the Microphysics}
The potential impact of uncertainties in the microphysics on the
supernova explosion mechanism has long been recognized.  We cannot
cover the full spectrum of ideas, which even includes less defensible
scenarios for achieving shock revival in spherical symmetry
\citep{fischer_11}. Instead we merely highlight some developments that
illustrate that modest (and therefore credible) variations of the
input physics could tip the balance in favour of shock revival in
multi-D models that are already close to the explosion threshold by
increasing $L_\nu$ and $E_\nu$.  This can be achieved, e.g., by
nuclear equations of state that are ``softer''
\citep{janka_12,suwa_13} in the sense that they result in a faster
contraction of the warm PNS and hence in hotter
neutrinospheres. Contrary to superficial
expectations, the increase of the neutrino luminosities
and mean energies outweighs the adverse effect
(increase of $|e_\mathrm{tot,g}|$) of the contraction
of the gain radius.

A similar beneficial effect on the heating conditions could result
from faster contraction of the PNS because of enhanced
heavy flavour neutrino losses. This was recently explored in 3D
simulations by \citet{melson_15b}, who found that a reduction of
neutral-current scattering opacities of the order of $\mathord{\sim} 20\%$ in the neutrinospheric region was sufficient to tip the scale
in favour of an explosion in a $20 M_\odot$ progenitor. \citet{melson_15b} achieved this opacity
reduction by assuming a stronger strangeness contribution
$g_\mathrm{a}^\mathrm{s}=-0.2$ to the axial vector coupling of the
nucleon than measured in current experiments. This is merely a
parameterization of nuclear physics uncertainties that can lead to
similar effects such as nucleon correlations; here the accuracy of
extant rate calculations \citep{burrows_98,reddy_99} based on the
random phase approximation (RPA) in the relevant density regime may be
questioned.  More reliable calculations of neutral-current scattering
opacities at neutrinospheric densities due to correlation effects are
now emerging \citep{horowitz_17} and being tested in supernova
simulations \citep{vartanyan_17}, but whether there are sizable
changes to extant models based on RPA opacities remains to be seen.

\begin{figure}[b]
\begin{center}
 \includegraphics[width=0.64 \textwidth]{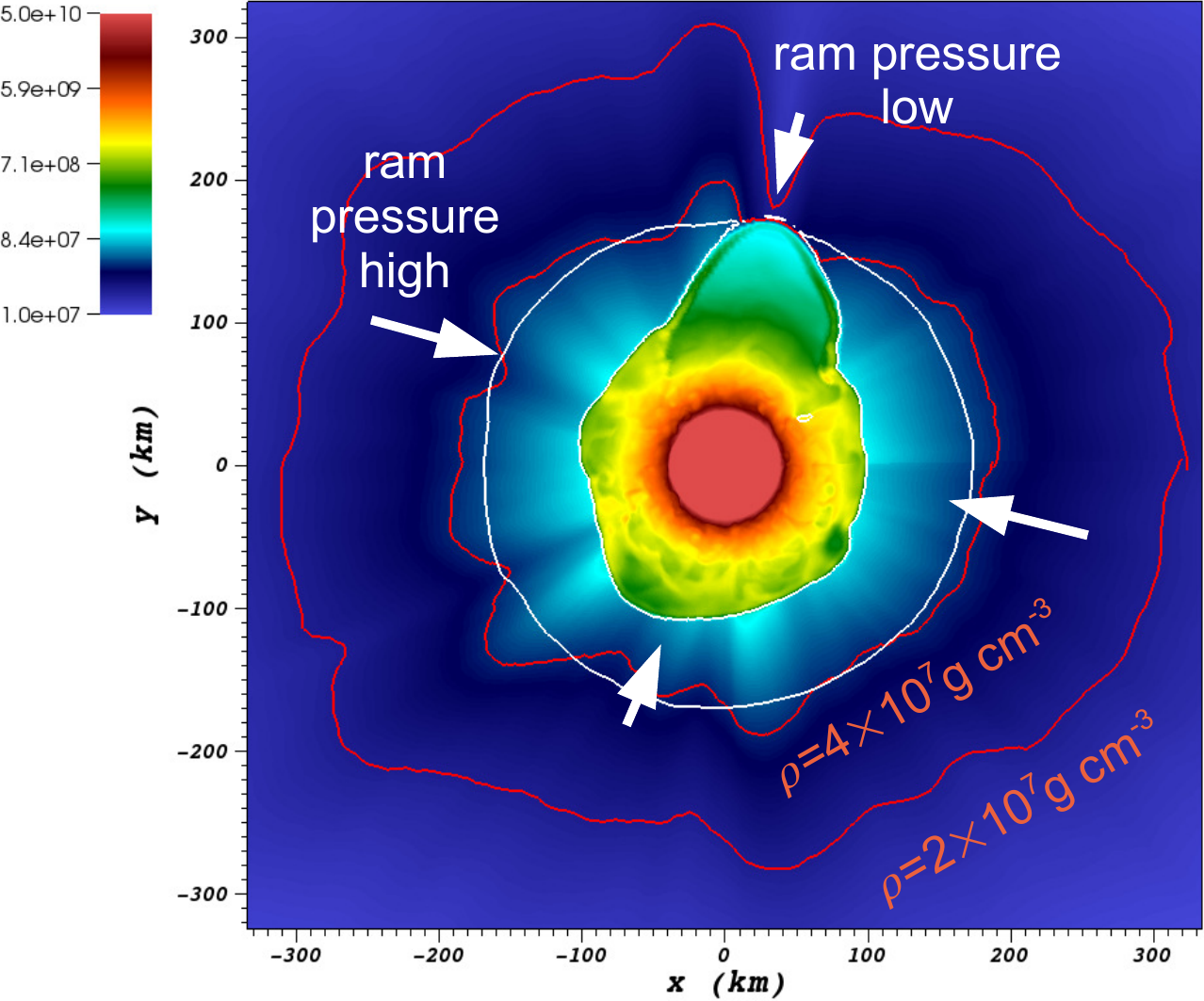}
 \caption{Illustration of the perturbation-aided version of the neutrino-driven mechanism,
showing a 2D slice of the density in a simulation of an $18 M_\odot$ star using
3D initial conditions from \citet{mueller_16b} at a time of
 $293\, \mathrm{ms}$ after bounce. The convective
velocity perturbations in the oxygen shell are
translated into density perturbations during the collapse
as can be seen by the distortion of density
isocontours (red), whereas variations
in the pre-shock radial velocity $v_r$ are small
(outer white curve: isocontour for $v_r=-4.5 \times 10^{9} \, \mathrm{cm}\, \mathrm{s}^{-1}$.
The variations in ram pressure significantly distort the shock
(inner white curve) and facilitate runaway
shock expansion by creating large and stable buoyant bubbles.
   \label{fig:perturbations}}
\end{center}
\end{figure}

\subsection{Perturbation-Aided Explosions and Uncertainties
in the Progenitor Structure} Another idea for more robust
neutrino-driven explosion models targets the term $\langle
\mathrm{Ma}^2\rangle$ in Equation~(\ref{eq:lcrit}) and seeks to identify effects
that increase the violence of convection and the SASI in order to
reduce the critical luminosity. This requires some additional
forcing for the instabilities in the gain region, which could arise
naturally due to the infall of shells with sizable seed asymmetries
from the late convective burning stages \citep{couch_13,mueller_15a}.
During the collapse, the initial velocity perturbations in such shells
give rise to density and pressure perturbations by advective-acoustic
coupling \citep{lai_00}. As illustrated
in Figure~\ref{fig:perturbations},
the anisotropic pre-shock ram pressure then
leads to increased shock deformation, the development of fast lateral
flows downstream of the shock as a result of shock obliquity \citep{mueller_15a},
and post-shock density perturbations that are translated into
turbulent kinetic energy by buoyancy \citep{mueller_16c}.

Various studies employing parameterised initial perturbations
\citep{couch_13,couch_14,mueller_15a,vartanyan_17} have been conducted
and established the conditions under which this ``perturbation-aided''
mechanism can be effective. Initial convective Mach numbers
$\mathord{\gtrsim} 0.05$ and large-scale flow with dominant
wavenumbers $\ell \lesssim 4$ are required for a significant reduction
of the critical luminosity \citep{mueller_15a}. More definitive
statements require supernova simulations starting from more
self-consistent 3D initial models. First attempts to evolve massive
stars through the last few minutes of convective shell burning before
collapse in 3D have been made by \citet{couch_15} for silicon burning
in a $15 M_\odot$ progenitor (assuming octant symmetry and a somewhat
problematic acceleration of the deleptonization of the iron core) and
by \citet{mueller_16c} for oxygen shell burning in an $18 M_\odot$
progenitor using a contracting inner boundary condition. Numerical
modelling of the final stages of massive stars is still in its infancy,
however. Silicon burning in particular presents a major technical
obstacle because it needs to be treated with larger nuclear reaction
networks than used in current 3D progenitor simulations. Moreover,
the 3D progenitors of \citet{couch_15} and \citet{mueller_16c}
have been computed under the proviso that the evolution up to
the last minutes before collapse is captured correctly by
spherically symmetric stellar evolution models (see below).

The first multi-group neutrino hydrodynamics simulation \citep{mueller_16b} based on the
3D progenitor model of \citet{mueller_16c} showed a strong dynamical
effect of the initial perturbations in the oxygen shell. Thanks to the
forced shock deformation, a neutrino-driven explosion develops $250 \,
\mathrm{ms}$ after bounce, whereas shock revival does not take place
in a control run with spherically symmetric initial conditions at
least for another $300 \, \mathrm{ms}$. By contrast,
the effect of perturbations was modest in the simpler leakage-based
simulations conducted by  \citet{couch_15} using their $15 M_\odot$
model. Despite the vastly different simulation methodology employed
by these two groups, this may already indicate that large
initial perturbations are not a panacea for the neutrino-driven
mechanism: 1D stellar evolution models in fact show a huge
spread of the parameters determining the efficiency
of the perturbation-aided mechanism. 
The convective Mach numbers in the relevant shells range between
$\mathord{\sim} 0.01$ and $\mathord{\sim} 0.15$ at the onset
of collapse, and the shell
thickness (which determines the dominant angular wavenumber $\ell$
of convective eddies) also varies tremendously. Pre-collapse
asphericities may therefore significantly help shock revival in
certain progenitors with violent shell burning, but play
a subdominant role in the explosion mechanism in others. 

While the idea of a perturbation-aided mechanism does not
require any \emph{fundamental} break with 1D stellar evolution models
based on mixing length theory (and merely needs an additional initialization
step before the onset of collapse), it has also been speculated that
the secular evolution of supernova progenitors may be more seriously
affected by multi-D processes such as turbulent entrainment
at convective boundaries \citep{meakin_07},
which might change $M$ and $\dot{M}$ in Equation~(\ref{eq:lcrit}). At this stage, it still appears
premature to draw conclusions based on 3D simulations of convective boundary
mixing \citep{meakin_07,mueller_16c,cristini_16,jones_17} during brief
intervals of advanced burning stages. More work is needed
to translate the findings from such multi-D models into suitable
recipes for 1D stellar evolution.

\subsection{Strong SASI, Rapid Rotation, and Magnetic Fields}
Several other mechanisms for facilitating neutrino-driven
shock revival have also been explored recently, although
it is less clear whether they could operate generically.
\citet{fernandez_15} found a \emph{reduction} of the critical
luminosity in the strongly SASI-dominated regime in 3D compared
to 2D in his light-bulb models, which he traced to the ability of
the non-axisymmetric SASI spiral mode to store
more non-radial kinetic energy than the axisymmetric
sloshing mode in the non-linear regime. Whether
and when this regime is realized remains to be further
investigated; the effect has not yet been
replicated by self-consistent simulations that probe
the SASI-dominated regime (e.g.\ the $27 M_\odot$
and $20 M_\odot$ models of \citealt{hanke_13}
and \citealt{melson_15b}), and may be restricted
to massive progenitors with sustained high accretion rates.

\citet{janka_16} and \citet{takiwaki_16} showed
that rapid rotation is conducive to shock revival
even without the help of magnetic fields. For increasing
rotation rate, this beneficial
effect initially comes about because the contribution
of rotational energy reduces the average net
binding energy  $|e_\mathrm{tot,g}|$ in the gain region,
and because angular momentum support lowers the
pre-shock velocity \citep{janka_16}. Above a critical
rotation rate corresponding to initial iron core
spin periods of $\mathord{\sim} 1\, \mathrm{s}$, the effect of rotation becomes
more dramatic as a strong corotation instability develops
\citep{takiwaki_16}.

There may also be a regime where magnetic fields play a subsidiary
role in the neutrino-driven mechanism instead of acting as the primary
driver of the explosion if the initial fields can be sufficiently
amplified by a small-scale turbulent dynamo, which could be provided
by convection \citep{thompson_93} or by the SASI
\citep{endeve_10,endeve_12} even in the absence of rapid rotation.
If the fields come close to equipartition strength,
they may help organize the flow into stable large-scale
bubbles and thereby prove conducive to shock revival \citep{obergaulinger_14}.
However,  strong initial fields of 
$\mathord{\sim} 10^{12} \, \mathrm{G}$ are still required
for this scenario in \citet{obergaulinger_14}

If either rotation or magnetic fields are to play at least a subsidiary role
in neutrino-driven explosions, it thus appears that 
special evolutionary channels are already required considering
that current stellar evolution models predict
pre-collapse spin periods of $\gtrsim 30\, \mathrm{s}$
and magnetic fields of $\lesssim 10^{10} \, \mathrm{G}$
\citep{heger_05} for solar-metallicity supernova progenitors.

\subsection{The Need for Alternative Mechanisms for Hypernovae}
That extremely fast rotation rates are encountered at least in a small
fraction of supernova progenitors is suggested by observations of
broad-lined Ic supernovae or ``hypernova'', whose explosion energies
reach up to $\mathord{\sim}10^{52} \, \mathrm{erg}$
\citep{drout_11}. These energies are likely out of reach for the
neutrino-driven mechanism and thus require a different explosion mechanism
altogether; even optimistic parameterised models suggest an upper
limit of $\mathord{\sim} 2\times 10^{51} \, \mathrm{erg}$ for 
neutrino-driven explosions. Various magnetohydrodynamic mechanisms 
\citep{uzov_92,macfadyen_99,akiyama_03} that tap
the rotational energy of a rapidly spinning and
highly magnetized ``millisecond magnetar'', black hole,
or accretion disk remain the most promising
explanations for the most energetic supernovae and the
gamma-ray-bursts associated with some of them, but we must
refer to the review of \citet{janka_12} for
a more detailed discussion.

\section{Outlook: From Neutrino-Driven Explosion
Models to Observables}
\label{sec:observables}
It is conceivable that 3D simulations of core-collapse supernovae may
need no more than a combination of relatively minor changes in the
input physics and slight improvements in numerical accuracy (for a 
discussion see \citealt{janka_16} and
\citealt{mueller_16b}) to produce neutrino-driven explosions over a
wide range of progenitor masses. This, however, would only provide a
solution to the problem of shock revival. The bigger challenge lies in accounting for the
observed explosion (e.g.\, explosion energy and nickel mass) and
remnant properties (neutron star mass, spin, and kick).

In neutrino-driven supernovae, the key explosion properties only reach
their asymptotic values after a phase of simultaneous accretion and
mass ejection that lasts for $\mathord{\gtrsim} 1 \, \mathrm{s}$ after shock
revival (or even later in the case of the neutron star kicks, see
\citealt{wongwathanarat_13}). Such time-scales are only marginally
within reach even for 3D hydrodynamics simulations with simplified
multi-group neutrino transport \citep{mueller_15b,mueller_16b}.
Axisymmetric models do not offer a serious alternative not only because of
their higher proclivity to explosion. 2D effects are even
more problematic during the explosion phase than prior to shock revival
\citep{mueller_15b} and may be responsible for the low explosion
energies (compared to typical observed values of $5\ldots 9 \times
10^{50} \, \mathrm{erg}$, see \citealp{kasen_09}) and high neutron
star masses found in typical 2D simulations
(\citealp{janka_12b,nakamura_15,oconnor_16}, but see also
the more energetic models of
\citealt{bruenn_16}). Because of these obstacles, it yet remains
to be demonstrated by self-consistent models
that the neutrino-driven mechanism can account for
the explosion properties of the majority of core-collapse supernovae
(and this statement holds \emph{a fortiori} for the magnetohydrodynamic
mechanism).

At present, the only alternative for understanding the systematics of
the explosion properties is to rely on parameterised models of
neutrino-driven supernovae
(e.g.\ \citealp{oconnor_10,ugliano_12,pejcha_15a,ertl_15,sukhbold_16,mueller_16a})
that are based on 1D simulations and/or analytic theory.  Although
objections may be raised against the predictiveness of such an
approach, it has undoubtedly proved valuable and led to significant
results.  Although the available parameterised models differ greatly
in detail, some findings such as trends in ``explodability'' (with
explosions up to $\mathord{\sim 15 M_\odot}$ and several islands of
explodability alternating with black hole formation up to
$\mathord{\sim 30 M_\odot}$), or the upper limit of $\mathord{\sim} 2
\times 10^{51} \, \mathrm{erg}$
\citep{ugliano_12,ertl_15,sukhbold_16,mueller_16a} have proved quite
robust, which suggests that they indeed reflect the inherent physics
of the neutrino-driven mechanism. Furthermore, first-principle
simulations are increasingly starting to inform parameterised models
\citep{mueller_16a}, and these in turn are proving useful for
selecting appropriate targets for full-scale multi-D
simulations. Given the prohibitive costs of state-of-the-art models,
such a two-pronged approach likely remains the best strategy for
explaining the diversity of core-collapse supernovae in the
foreseeable future.

\begin{acknowledgements}
The author acknowledges long-term assistance by his collaborators,
especially A.~Heger, H.-Th.~Janka, and T.~Melson, and support by the
STFC DiRAC HPC Facility (DiRAC Data Centric system, ICC Durham), the
National Computational Infrastructure (Australia), the Pawsey
Supercomputing Centre (University of Western Australia), and
the Minnesota Supercomputing Institute.

\end{acknowledgements}


\begin{thebibliography}{61}
\expandafter\ifx\csname natexlab\endcsname\relax\def\natexlab#1{#1}\fi

\bibitem[{{Akiyama} {et~al.}(2003){Akiyama}, {Wheeler}, {Meier},
  {Lichtenstadt}, \& {Lichtenstadt}}]{akiyama_03}
{Akiyama}, S., {Wheeler}, J.~C., {Meier}, D.~L., {Lichtenstadt}, I.~{Meier},
  D.~L., \& {Lichtenstadt}. 2003, \apj, 584, 954

\bibitem[{{Blondin} {et~al.}(2003){Blondin}, {Mezzacappa}, \&
  {DeMarino}}]{blondin_03}
{Blondin}, J.~M., {Mezzacappa}, A., \& {DeMarino}, C. 2003, \apj, 584, 971

\bibitem[{{Bruenn} {et~al.}(2016){Bruenn}, {Lentz}, {Hix}, {Mezzacappa},
  {Harris}, {Messer}, {Endeve}, {Blondin}, {Chertkow}, {Lingerfelt},
  {Marronetti}, \& {Yakunin}}]{bruenn_16}
{Bruenn}, S.~W. {et~al.} 2016, \apj, 818, 123

\bibitem[{{Bruenn} {et~al.}(2013){Bruenn}, {Mezzacappa}, {Hix}, {Lentz},
  {Bronson Messer}, {Lingerfelt}, {Blondin}, {Endeve}, {Marronetti}, \&
  {Yakunin}}]{bruenn_13}
{Bruenn}, S.~W. {et~al.} 2013, \apjl, 767, L6

\bibitem[{{Burrows} \& {Goshy}(1993)}]{burrows_93}
{Burrows}, A., \& {Goshy}, J. 1993, \apjl, 416, L75+

\bibitem[{{Burrows} \& {Sawyer}(1998)}]{burrows_98}
{Burrows}, A., \& {Sawyer}, R.~F. 1998, \prc, 58, 554

\bibitem[{{Burrows} {et~al.}(2016){Burrows}, {Vartanyan}, {Dolence}, {Skinner},
  \& {Radice}}]{vartanyan_17}
{Burrows}, A., {Vartanyan}, D., {Dolence}, J.~C., {Skinner}, M.~A., \&
  {Radice}, D. 2016, ArXiv e-prints, 1611.05859

\bibitem[{{Cantiello} {et~al.}(2014){Cantiello}, {Mankovich}, {Bildsten},
  {Christensen-Dalsgaard}, \& {Paxton}}]{cantiello_14}
{Cantiello}, M., {Mankovich}, C., {Bildsten}, L., {Christensen-Dalsgaard}, J.,
  \& {Paxton}, B. 2014, \apj, 788, 93

\bibitem[{{Couch}(2013)}]{couch_12b}
{Couch}, S.~M. 2013, \apj, 775, 35

\bibitem[{{Couch} {et~al.}(2015){Couch}, {Chatzopoulos}, {Arnett}, \&
  {Timmes}}]{couch_15}
{Couch}, S.~M., {Chatzopoulos}, E., {Arnett}, W.~D., \& {Timmes}, F.~X. 2015,
  \apjl, 808, L21

\bibitem[{{Couch} \& {Ott}(2013)}]{couch_13}
{Couch}, S.~M., \& {Ott}, C.~D. 2013, \apjl, 778, L7

\bibitem[{{Couch} \& {Ott}(2015)}]{couch_14}
{Couch}, S.~M.. 2015, \apj, 799, 5

\bibitem[{{Cristini} {et~al.}(2016){Cristini}, {Meakin}, {Hirschi}, {Arnett},
  {Georgy}, \& {Viallet}}]{cristini_16}
{Cristini}, A., {Meakin}, C., {Hirschi}, R., {Arnett}, D., {Georgy}, C., \&
  {Viallet}, M. 2016, \physscr, 91, 034006

\bibitem[{{Drout} {et~al.}(2011){Drout}, {Soderberg}, {Gal-Yam}, {Cenko},
  {Fox}, {Leonard}, {Sand}, {Moon}, {Arcavi}, \& {Green}}]{drout_11}
{Drout}, M.~R. {et~al.} 2011, \apj, 741, 97

\bibitem[{{Endeve} {et~al.}(2010){Endeve}, {Cardall}, {Budiardja}, \&
  {Mezzacappa}}]{endeve_10}
{Endeve}, E., {Cardall}, C.~Y., {Budiardja}, \& {Mezzacappa}, A. 2010, \apj,
  713, 1219

\bibitem[{{Endeve} {et~al.}(2012){Endeve}, {Cardall}, {Budiardja}, {Beck},
  {Bejnood}, {Toedte}, {Mezzacappa}, \& {Blondin}}]{endeve_12}
{Endeve}, E., {Cardall}, C.~Y., {Budiardja}, R.~D., {Beck}, S.~W., {Bejnood},
  A., {Toedte}, R.~J., {Mezzacappa}, A., \& {Blondin}, J.~M. 2012, \apj, 751,
  26

\bibitem[{{Ertl} {et~al.}(2016){Ertl}, {Janka}, {Woosley}, {Sukhbold}, \&
  {Ugliano}}]{ertl_15}
{Ertl}, T., {Janka}, H.-T., {Woosley}, S.~E., {Sukhbold}, T., \& {Ugliano}, M.
  2016, \apj, 818, 124

\bibitem[{{Fern{\'a}ndez}(2015)}]{fernandez_15}
{Fern{\'a}ndez}, R. 2015, \mnras, 452, 2071

\bibitem[{{Fischer} {et~al.}(2011){Fischer}, {Sagert}, {Pagliara}, {Hempel},
  {Schaffner-Bielich}, {Rauscher}, {Thielemann}, {K{\"a}ppeli},
  {Mart{\'{\i}}nez-Pinedo}, \& {Liebend{\"o}rfer}}]{fischer_11}
{Fischer}, T. {et~al.} 2011, \apjs, 194, 39

\bibitem[{{Hanke} {et~al.}(2012){Hanke}, {Marek}, {M{\"u}ller}, \&
  {Janka}}]{hanke_12}
{Hanke}, F., {Marek}, A., {M{\"u}ller}, B., \& {Janka}, H.-T. 2012, \apj, 755,
  138

\bibitem[{{Hanke} {et~al.}(2013){Hanke}, {M{\"u}ller}, {Wongwathanarat},
  {Marek}, \& {Janka}}]{hanke_13}
{Hanke}, F., {M{\"u}ller}, B., {Wongwathanarat}, A., {Marek}, A., \& {Janka},
  H.-T. 2013, \apj, 770, 66

\bibitem[{{Heger} {et~al.}(2005){Heger}, {Woosley}, \& {Spruit}}]{heger_05}
{Heger}, A., {Woosley}, S.~E., \& {Spruit}, H.~C. 2005, \apj, 626, 350

\bibitem[{{Herant} {et~al.}(1994){Herant}, {Benz}, {Hix}, {Fryer}, \&
  {Colgate}}]{herant_94}
{Herant}, M., {Benz}, W., {Hix}, W.~R., {Fryer}, C.~L., \& {Colgate}, S.~A.
  1994, \apj, 435, 339

\bibitem[{{Horowitz} {et~al.}(2016){Horowitz}, {Caballero}, {Lin}, {O'Connor},
  \& {Schwenk}}]{horowitz_17}
{Horowitz}, C.~J., {Caballero}, O.~L., {Lin}, Z., {O'Connor}, E., \& {Schwenk},
  A. 2016, ArXiv e-prints, 1611.05140

\bibitem[{{Janka}(2012)}]{janka_12}
{Janka}, H.-T. 2012, Annual Review of Nuclear and Particle Science, 62, 407

\bibitem[{{Janka} {et~al.}(2012){Janka}, {Hanke}, {H{\"u}depohl}, {Marek},
  {M{\"u}ller}, \& {Obergaulinger}}]{janka_12b}
{Janka}, H.-T., {Hanke}, F., {H{\"u}depohl}, L., {Marek}, A., {M{\"u}ller}, B.,
  \& {Obergaulinger}, M. 2012, Progress of Theoretical and Experimental
  Physics, 2012, 010000

\bibitem[{{Janka} {et~al.}(2016){Janka}, {Melson}, \& {Summa}}]{janka_16}
{Janka}, H.-T., {Melson}, T., \& {Summa}, A. 2016, Annual Review of Nuclear and
  Particle Science, 66, 341, 1602.05576

\bibitem[{{Jones} {et~al.}(2017){Jones}, {Andrassy}, {Sandalski}, {Davis},
  {Woodward}, \& {Herwig}}]{jones_17}
{Jones}, S., {Andrassy}, R., {Sandalski}, S., {Davis}, A., {Woodward}, P., \&
  {Herwig}, F. 2017, \mnras, 465, 2991

\bibitem[{{Kasen} \& {Woosley}(2009)}]{kasen_09}
{Kasen}, D., \& {Woosley}, S.~E. 2009, \apj, 703, 2205

\bibitem[{{Kitaura} {et~al.}(2006){Kitaura}, {Janka}, \&
  {Hillebrandt}}]{kitaura_06}
{Kitaura}, F.~S., {Janka}, H.-T., \& {Hillebrandt}, W. 2006, \aap, 450, 345

\bibitem[{{Lai} \& {Goldreich}(2000)}]{lai_00}
{Lai}, D., \& {Goldreich}, P. 2000, \apj, 535, 402

\bibitem[{{Lentz} {et~al.}(2015){Lentz}, {Bruenn}, {Hix}, {Mezzacappa},
  {Messer}, {Endeve}, {Blondin}, {Harris}, {Marronetti}, \&
  {Yakunin}}]{lentz_15}
{Lentz}, E.~J. {et~al.} 2015, \apjl, 807, L31

\bibitem[{{MacFadyen} \& {Woosley}(1999)}]{macfadyen_99}
{MacFadyen}, A.~I., \& {Woosley}, S.~E. 1999, \apj, 524, 262

\bibitem[{{Meakin} \& {Arnett}(2007)}]{meakin_07}
{Meakin}, C.~A., \& {Arnett}, D. 2007, \apj, 667, 448

\bibitem[{{Melson} {et~al.}(2015{\natexlab{a}}){Melson}, {Janka}, {Bollig},
  {Hanke}, {Marek}, \& {M{\"u}ller}}]{melson_15b}
{Melson}, T., {Janka}, H.-T., {Bollig}, R., {Hanke}, F., {Marek}, A., \&
  {M{\"u}ller}, B. 2015{\natexlab{a}}, \apjl, 808, L42

\bibitem[{{Melson} {et~al.}(2015{\natexlab{b}}){Melson}, {Janka}, \&
  {Marek}}]{melson_15a}
{Melson}, T., {Janka}, H.-T., \& {Marek}, A. 2015{\natexlab{b}}, \apjl, 801,
  L24

\bibitem[{{M{\"u}ller}(2015)}]{mueller_15b}
{M{\"u}ller}, B. 2015, \mnras, 453, 287

\bibitem[{{M{\"u}ller}(2016)}]{mueller_16b}
{M{\"u}ller}, B. 2016, \pasa, 33, e048

\bibitem[{{M{\"u}ller} {et~al.}(2016{\natexlab{a}}){M{\"u}ller}, {Heger},
  {Liptai}, \& {Cameron}}]{mueller_16a}
{M{\"u}ller}, B., {Heger}, A., {Liptai}, D., \& {Cameron}, J.~B.
  2016{\natexlab{a}}, \mnras, 460, 742

\bibitem[{{M{\"u}ller} \& {Janka}(2015)}]{mueller_15a}
{M{\"u}ller}, B., \& {Janka}, H.-T. 2015, \mnras, 448, 2141

\bibitem[{{M{\"u}ller} {et~al.}(2016{\natexlab{b}}){M{\"u}ller}, {Viallet},
  {Heger}, \& {Janka}}]{mueller_16c}
{M{\"u}ller}, B., {Viallet}, M., {Heger}, A., \& {Janka}, H.-T.
  2016{\natexlab{b}}, \apj, 833, 124

\bibitem[{{Murphy} {et~al.}(2013){Murphy}, {Dolence}, \& {Burrows}}]{murphy_12}
{Murphy}, J.~W., {Dolence}, J.~C., \& {Burrows}, A. 2013, \apj, 771, 52

\bibitem[{{Nakamura} {et~al.}(2014){Nakamura}, {Kuroda}, {Takiwaki}, \&
  {Kotake}}]{nakamura_14}
{Nakamura}, K., {Kuroda}, T., {Takiwaki}, T., \& {Kotake}, K. 2014, \apj, 793,
  45

\bibitem[{{Nakamura} {et~al.}(2015){Nakamura}, {Takiwaki}, {Kuroda}, \&
  {Kotake}}]{nakamura_15}
{Nakamura}, K., {Takiwaki}, T., {Kuroda}, T., \& {Kotake}, K. 2015, \pasj, 67,
  107

\bibitem[{{Nordhaus} {et~al.}(2010){Nordhaus}, {Burrows}, {Almgren}, \&
  {Bell}}]{nordhaus_10}
{Nordhaus}, J., {Burrows}, A., {Almgren}, A., \& {Bell}, J. 2010, \apj, 720,
  694

\bibitem[{{Obergaulinger} {et~al.}(2014){Obergaulinger}, {Janka}, \&
  {Aloy}}]{obergaulinger_14}
{Obergaulinger}, M., {Janka}, H.-T., \& {Aloy}, M.~A. 2014, \mnras, 445, 3169

\bibitem[{{O'Connor} \& {Couch}(2015)}]{oconnor_16}
{O'Connor}, E., \& {Couch}, S. 2015, ArXiv e-prints, 1511.07443

\bibitem[{{O'Connor} \& {Ott}(2010)}]{oconnor_10}
{O'Connor}, E., \& {Ott}, C.~D. 2010, Classical and Quantum Gravity, 27, 114103

\bibitem[{{Pejcha} \& {Thompson}(2015)}]{pejcha_15a}
{Pejcha}, O., \& {Thompson}, T.~A. 2015, \apj, 801, 90

\bibitem[{{Reddy} {et~al.}(1999){Reddy}, {Prakash}, {Lattimer}, \&
  {Pons}}]{reddy_99}
{Reddy}, S., {Prakash}, M., {Lattimer}, J.~M., \& {Pons}, J.~A. 1999, \prc, 59,
  2888

\bibitem[{{Smartt}(2015)}]{smartt_15}
{Smartt}, S.~J. 2015, \pasa, 32, 16

\bibitem[{{Sukhbold} {et~al.}(2016){Sukhbold}, {Ertl}, {Woosley}, {Brown}, \&
  {Janka}}]{sukhbold_16}
{Sukhbold}, T., {Ertl}, T., {Woosley}, S.~E., {Brown}, J.~M., \& {Janka}, H.-T.
  2016, \apj, 821, 38

\bibitem[{{Summa} {et~al.}(2016){Summa}, {Hanke}, {Janka}, {Melson}, {Marek},
  \& {M{\"u}ller}}]{summa_16}
{Summa}, A., {Hanke}, F., {Janka}, H.-T., {Melson}, T., {Marek}, A., \&
  {M{\"u}ller}, B. 2016, \apj, 825, 6, 1511.07871

\bibitem[{{Suwa} {et~al.}(2013){Suwa}, {Takiwaki}, {Kotake}, {Fischer},
  {Liebend{\"o}rfer}, \& {Sato}}]{suwa_13}
{Suwa}, Y., {Takiwaki}, T., {Kotake}, K., {Fischer}, T., {Liebend{\"o}rfer},
  M., \& {Sato}, K. 2013, \apj, 764, 99

\bibitem[{{Takiwaki} {et~al.}(2014){Takiwaki}, {Kotake}, \&
  {Suwa}}]{takiwaki_14}
{Takiwaki}, T., {Kotake}, K., \& {Suwa}, Y. 2014, \apj, 786, 83

\bibitem[{{Takiwaki} {et~al.}(2016){Takiwaki}, {Kotake}, \&
  {Suwa}}]{takiwaki_16}
{Takiwaki}, T., {Kotake}, K., \& {Suwa}, Y. 2016, \mnras, 461, L112

\bibitem[{{Tamborra} {et~al.}(2014){Tamborra}, {Hanke}, {Janka}, {M{\"u}ller},
  {Raffelt}, \& {Marek}}]{tamborra_14a}
{Tamborra}, I., {Hanke}, F., {Janka}, H.-T., {M{\"u}ller}, B., {Raffelt},
  G.~G., \& {Marek}, A. 2014, \apj, 792, 96

\bibitem[{{Thompson} \& {Duncan}(1993)}]{thompson_93}
{Thompson}, C., \& {Duncan}, R.~C. 1993, \apj, 408, 194

\bibitem[{{Ugliano} {et~al.}(2012){Ugliano}, {Janka}, {Marek}, \&
  {Arcones}}]{ugliano_12}
{Ugliano}, M., {Janka}, H.-T., {Marek}, A., \& {Arcones}, A. 2012, \apj, 757,
  69

\bibitem[{{Usov}(1992)}]{uzov_92}
{Usov}, V.~V. 1992, \nat, 357, 472

\bibitem[{{Wongwathanarat} {et~al.}(2013){Wongwathanarat}, {Janka}, \&
  {M{\"u}ller}}]{wongwathanarat_13}
{Wongwathanarat}, A., {Janka}, H.-T., \& {M{\"u}ller}, E. 2013, \aap, 552, A126

\end{thebibliography}
\end{document}